\documentclass[aps,pra,preprint,superscriptaddress,showpacs]{revtex4-1}
\usepackage{amsmath,amsfonts}
\usepackage[colorlinks,linkcolor=blue, urlcolor=blue, anchorcolor=blue, citecolor=blue]{hyperref}
\usepackage[T1]{fontenc}
\usepackage{mathptmx}
\usepackage{graphicx}
\usepackage{amsmath}
\usepackage{amssymb}
\DeclareMathAlphabet{\mathcal}{OMS}{cmsy}{m}{n}  %mathptmx \mathcal{}

\begin{document}

%\preprint{}

\title{Optimal quantum phase estimation in an atomic gyroscope based on Bose-Hubbard model}


\author{Lei Shao}
\author{Weiyao Li}
\author{Xiaoguang Wang}
 \email{xgwang1208@zju.edu.cn}
%\thanks{}
\affiliation{Zhejiang Institute of Modern Physics, Department of Physics, Zhejiang University, Hangzhou 310027, China}


\begin{abstract}
We investigate the optimal quantum state for an atomic gyroscope based on a three-site Bose-Hubbard model. In  previous studies, various states such as the uncorrelated state, the BAT state and the NOON state are employed as the probe states to estimate the phase uncertainty. In this article, we present a Hermitian operator $\mathcal{H}$ and an equivalent unitary parametrization transformation to calculate the quantum Fisher information for any initial states. Exploiting this equivalent unitary parametrization transformation, we can seek the optimal state which gives the maximal quantum Fisher information on both lossless and lossy conditions. As a result, we find that the entangled even squeezed state (EESS) can significantly enhance the precision for moderate loss rates.
\end{abstract}

%\keywords{}

\maketitle

% References should be done using the~\cite, \ref, and \label commands
\section{\label{sec:level1}Introduction}
In recent decades, the technology of quantum physics plays an important role in precision measurements and the accuracy has been improved significantly in comparison with  classical systems. The advantages of quantum technology are reflected in various fields such as  the biological sensing~\cite{TAYLOR20161,Taylor2013,NASR20091154,Morris2015}, the measurements of physical constants~\cite{Fixler74,PhysRevA.88.043610,PhysRevLett.100.050801}, and the gravitational wave detection~\cite{PhysRevA.88.041802,Purdy801,Punturo_2010,BRAGINSKY2004345,RevModPhys.86.121,PhysRevA.65.033608}. In quantum metrology, the optical interferometer such as the Mach-Zehnder (MZ) interferometer is frequently used to estimate the relative phase of two modes, and the precision obtained by using classic states can reach the standard quantum limit (SQL), i.e., $1/\sqrt{N}$ , where $N$ is the total mean number of photons of two modes. In 1981, a typical scheme proposed by Caves~\cite{PhysRevD.23.1693} is to take a coherent state $\left|\alpha\right\rangle$ and a squeezed vacuum state $\left|\xi\right\rangle$ as the input states of the Mach-Zehnder interferometer, which can beat the SQL. On the other hand, the unique characteristics of quantum states such as entanglement provide us with a way to reach the Heisenberg limit, i.e., $1/N$, and the NOON state~\cite{PhysRevLett.102.040403,doi:10.1080/0950034021000011536,PhysRevA.81.043624,Tsarev:18} is widely studied since its quantum property of maximal entanglement and superior performance in metrology. Based on the result of the NOON state, the entangled coherent state (ECS)~\cite{PhysRevLett.107.083601,PhysRevA.86.043828,PhysRevA.88.043832,Liu_2016} which is viewed as a similar probe state, gives a remarkable precision enhancement on both lossless and lossy conditions than the NOON state.

Similar researches for phase estimation can be applied to fiber-optic gyroscopes \cite{Arditty:81} and atomic gyroscopes, which are designed to measure the phase caused by rotation and broadly used for navigation systems. Atomic gyroscopes have received considerable attention for their excellent performance in enhancing sensitivity of rotation, and atomic Sagnac interferometers such as light-pulse atomic interferometers \cite{PhysRevLett.78.2046,Kasevich1992,Gustavson_2000},  and guided matter-wave interferometers \cite{PhysRevLett.99.173201,GarridoAlzar:12,PhysRevLett.115.163001,Rico_Gutierrez_2015} are widely studied. Another novel type of gyroscope is based on Bose-Hubbard model and it is composed of a collection of ultracold atoms trapped in an optical lattice loop of several sites~\cite{Cooper_2009,PhysRevA.81.043624}. In Ref.~\cite{PhysRevA.81.043624}, Cooper et al. investigated three different input states and found that the NOON state produces the best precision with scaling $1/N$ under lossless conditions. However, considering the presence of particle loss in practical systems, the BAT state is a better choice because of its robustness in high loss regime. In the above studies, quantum Fisher information (QFI)~\cite{Helstrom1976,PhysRevLett.72.3439,Holevo1982,BRAUNSTEIN1996135,Liu_2019} is an important concept in quantum metrology, which gives the lower limit of the variance of parameter $\theta$ due to the quantum Cram\'er-Rao bound, i.e.,
\begin{equation}\label{E1}
\Delta\theta\ge\frac{1}{\sqrt{\mu F_{Q}}},
\end{equation}
where $\mu$ is the number of independent repeats of the experiment and $F_{Q}$ is the QFI. Hence, the core task in many studies is trying every means to obtain larger QFI. For a general measurement scheme, quantum states passing through an atomic gyroscope or optical devices will cause a phase $\theta$, and such a variation can be regarded as a unitary parametrization processes $U(\theta)$. For a pure initial state $\left|\psi\right\rangle_{\rm in}$, the QFI with respect to the parameter $\theta$ is defined as \cite{PhysRevA.90.022117,Liu2015,PhysRevA.92.012312}
\begin{equation}\label{E2}
F_{Q}=4\left(\left\langle \mathcal{H}^{2}\right\rangle -\left\langle \mathcal{H}\right\rangle ^{2}\right),
\end{equation}
where
\begin{equation}\label{E3}
\mathcal{H}\equiv i\left(\partial_{\theta}U^{\dagger}\right)U
\end{equation}is a Hermitian operator and the QFI is determined by the variance of $\mathcal{H}$.  Thus the dynamical property of a definite unitary parametrization process is characterized by a fixed $\mathcal{H}$, and the optimal probe state could be gotten by comparing various input states.

In this article, we follow the work of the Ref.~\cite{PhysRevA.81.043624}, and propose an equivalent unitary parametrization processes $U_{\rm eq}$ to calculate the highest precision we can reach with the atomic gyroscope introduced above. Although three input states, i.e., the uncorrelated state, the BAT state and the NOON state have been discussed before, there are still superior states can be chosen to boost the precision. We also investigate the effects of  particle loss in the atomic gyroscope. In principle, the maximally entangled states such as the NOON state are vulnerable, and the phase information will be rapidly lost when decoherence occurs~\cite{PhysRevA.81.043624,PhysRevLett.107.083601,Tsarev:18,PhysRevA.78.063828}. Therefore, it is important to find a state that can reach the Heisenberg limit and meanwhile has good robustness against decoherence. Apart from the ECS involved in Ref.~\cite{PhysRevLett.107.083601}, the squeezed entangled state (SES)~\cite{PhysRevA.93.033859,Lee2016,Rubio_2019} which can be regarded as the superposition of NOON states with different even particle numbers shows great potential. The comparison between the ECS and the SES is similar to the relationship between the uncorrelated state and the BAT state discussed in Ref.~\cite{PhysRevA.81.043624}, which shows the superposition of even number states produces better precision and robustness than the superposition of the general number states. Furthermore, Refs.~\cite{Lee2016,PhysRevA.93.033859} indicate that the Mandel $\mathcal{Q}$-parameter of the single-mode in a path-symmetric state determines the upper limit of precision. Inspired by these work , we seek for a state can give larger Mandel-$\mathcal{Q}$ parameter, and this leads us to  propose "entangled even squeezed state" (EESS). The numerical simulation show that the EESS can give a significant precision enhancement in the small-particle-number regime, and this advantage is maintained until particle loss is larger than $50\%$. 

This article is organized as follow. In Section \ref{sec:level2}, we  give a Hermitian operator and general expression to calculate the QFI of atomic gyroscope. In In Section \ref{sec:level3}, we discuss the influence of different initial state on the precision in lossless case.  In Section \ref{sec:level8}, we consider the effects of particle loss on  the atomic gyroscope.
\section{\label{sec:level2}GENERAL EXPRESSION OF QFI}
\begin{figure}
	\includegraphics[width=98mm,height=32mm]{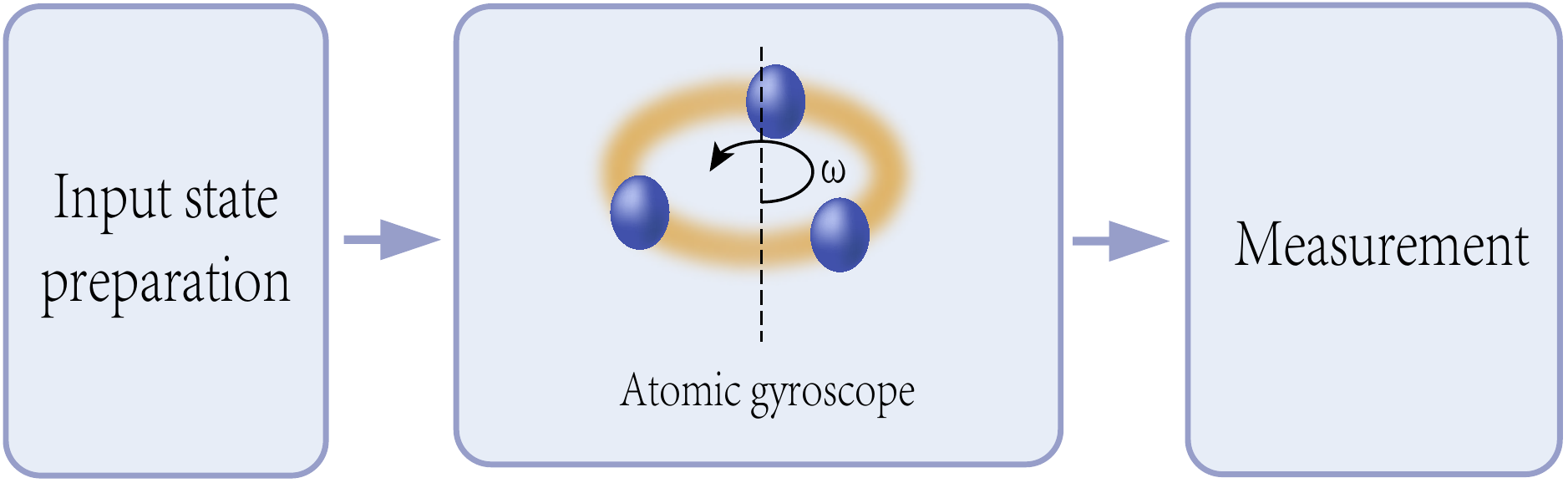}
	\centering% Here is how to import EPS art
	\caption{\label{fig:1} Schematic diagram of the phase estimation in the atomic gyroscope. A quantum state is prepared as an input state of the atomic gyroscope which is a ring configuration of three sites. Then the mechanical rotation of the atomic gyroscope causes a phase $\theta$, and after performing several operations the phase $\theta$ is able to read out.
	}
\end{figure}

First of all, it is necessary to review the procedure of phase measurement in Ref.~\cite{PhysRevA.81.043624}. The atomic gyroscope is composed of $N$ ultracold atoms of mass $m$ trapped in an optical lattice loop of three sites where the circumference of the loop is $L$ \cite{doi:10.1080/09500340701427128,Cooper_2009,doi:10.1063/1.4746058}, and a schematic diagram is shown in Fig. \ref{fig:1}. This scheme can be described by Bose-Hubbard Hamiltonian
\begin{equation}\label{E4}
\frac{H}{\hbar}=\sum_{i=0}^{2}\epsilon_{i}a_{i}^{\dagger}a_{i}-\sum_{i=0}^{2}J_{i}\left(a_{i}^{\dagger}a_{i+1}+a_{i+1}^{\dagger}a_{i}\right)+\sum_{i=0}^{2}V_{i}a_{i}^{\dagger2}a_{i}^{2},
\end{equation}
where $a_{i}^{\dagger}$ and $a_{i}$ are creation and annihilation operators in site $i$. The first term accounts for energy offset, and it can be ignored by taking a fixed zero-energy for each site, and we set $\epsilon_{i}=0$. The second term is the coupling between site $i$ and site $i+1$ , and the last term represents the interaction between atoms on each site. Moreover, the strengths of coupling energy and interaction energy are described by the parameters $J_{i}$ and $V_{i}$ respectively, and the latter can be ignored compared with the former when the potential barrier between two sites is reduced, i.e., $V_{i}\approx0$. This system is described in detail in Ref.~\cite{Cooper_2009}. Therefore, the unitary transformation $U=e^{-iHt/\hbar}$ only  depends on the coupling energy in the high coupling regime. It is convenient to diagonalize the Hamiltonian by using the quasi-momentum basis, or flow basis, \cite{doi:10.1080/09500340701427128,doi:10.1063/1.4746058,Hallwood_2006}, i.e.,
\begin{equation}\label{E5}
\begin{pmatrix}
\alpha_{-\!\!1}\\\alpha_{\,0}\\\alpha_{\,1}
\end{pmatrix}
=\frac{1}{\sqrt{3}}
\begin{pmatrix}
1&e^{-i2\pi/3}&e^{i2\pi/3}\\
1&1&1\\
1&e^{i2\pi/3}&e^{-i2\pi/3}
\end{pmatrix}
\begin{pmatrix}
a_{0}\\a_{1}\\a_{2}
\end{pmatrix}
\end{equation}
and the Hamiltonian in high coupling regime is given by
\begin{equation}\label{E6}
\frac{H}{\hbar}=-2J\sum_{j=-1}^{1}\cos\left(2\pi j/3\right)\alpha_{j}^{\dagger}\alpha_{j}.
\end{equation}
In this way, the three-port beam splitter (tritter) described in Ref.~\cite{Cooper_2009} is able to be realized. Specifically, the unitary transformation of the tritter is expressed as

\begin{align}\label{E7}
U_{2}&=e^{i\frac{2\pi}{9}\left(a_{0}^{\dagger}a_{1}+a_{1}^{\dagger}a_{2}+a_{2}^{\dagger}a_{0}+h.c.\right)}\nonumber\\
&=e^{i\frac{2\pi}{9}\left(2\alpha_{0}^{\dagger}\alpha_{0}-\alpha_{1}^{\dagger}\alpha_{1}-\alpha_{-1}^{\dagger}\alpha_{-1}\right)}
\end{align}

The inverse tritter operation is realized by changing the phase from $2\pi/9$ to $4\pi/9$. The procedure of phase estimation can be briefly summarized as follows: (\romannumeral1) Prepare an initial state. (\romannumeral2) Perform a tritter operation. (\romannumeral3) Apply a $2\pi/3$ phase to site two with the unitary transformation $U_{3}=\exp(i2\pi a^{\dagger}_{2}a_{2}/3)$. (\romannumeral4) Rotate the system with velocity $\omega$. (\romannumeral5) Apply a $-2\pi/3$ phase to site two with the unitary transformation $U_{5}=\exp(-i2\pi a^{\dagger}_{2}a_{2}/3)$. (\romannumeral6) Perform an inverse tritter operation. (\romannumeral7) Read out the phase $\theta$ caused by the rotation. The Hamiltonian used in step (\romannumeral4) is
\cite{PhysRevA.81.043624,Cooper_2009}
\begin{equation}\label{E8}
\frac{H_{r}}{\hbar}=-2J\sum_{j=-1}^{1}\cos\left(\theta/3-2\pi j/3\right)\alpha_{j}^{\dagger}\alpha_{j}.
\end{equation}

The purpose of step (\romannumeral2) and step (\romannumeral3) is to realize the transformation from the particle number states of each site to the flow states, i.e., $\left|l,m,n\right\rangle_{a_{0},a_{1},a_{2}}\Rightarrow$ $\left|l,m,n\right\rangle_{\alpha_{-1},\alpha_{0},\alpha_{1}}$. The term on the left side represents the number of atoms in each site, and the term on the right side represents the number of atoms in the $\alpha_{-1}$,$\alpha_{0}$ and $\alpha_{1}$ flow states, respectively. Similarly, step (\romannumeral5) and step (\romannumeral6) are the inverse operations of step (\romannumeral2) and step (\romannumeral3). One can immediately see that the Hamiltonian in Eq.~(\ref{E8}) is diagonalized and can act on the flow states directly. More details of this procedure are described in Ref.~\cite{PhysRevA.81.043624}.

To obtain the velocity of rotation more accurately, we need to reduce the variance of $\theta$ since $\Delta\omega=\Delta\theta\cdot\left(h/L^{2}m\right)$. It is worthy to note that the QFI for various initial states only depends on step (\romannumeral1) to (\romannumeral4), hence the ultimate precision of the atomic gyroscope is able to be calculated by a general Hermitian operator $\mathcal{H}$, which can be expressed as

\begin{equation}\label{E9}
\begin{split}
\mathcal{H}&=U_{2}^{\dagger}U_{3}^{\dagger}\mathcal{H}_{4}U_{3}U_{2}\\
&=\frac{-2Jt_{\omega}}{3}\left[\sin\left(\frac{\theta+2\pi}{3}\right)n_{0}+\sin\left(\frac{\theta-2\pi}{3}\right)n_{1}+\sin\left(\frac{\theta}{3}\right)n_{2}\right],
\end{split}
\end{equation}
where $\mathcal{H}_{4}=i\left(\partial_{\theta}U_{4}^{\dagger}\right)U_{4}=-\partial_{\theta}H_{r}t_{\omega}/\hbar$ according to Eq.~(\ref{E3}) and $U_{4}=e^{-iH_{r}t_{\omega}/\hbar}$  is the unitary transformation of rotation, and $t_{\omega}$ is the time the gyroscope takes to rotate. The detailed derivation of $\mathcal{H}$ is presented in Appendix \ref{Appendix:A}. Utilizing Eqs.~(\ref{E2}) and (\ref{E9}), a general expression of QFI for the atomic gyroscope is obtained, which is helpful to find the optimal state corresponds to this scheme.

In addition, the overall process of phase estimation can be simplified as an equivalent unitary parametrization transformation with Eqs.~(\ref{E3}) and (\ref{E9}),

\begin{equation}\label{E10}
\begin{split}
U_{\rm eq}&=\exp\left(i\phi_{\,0}n\right)\exp\left(i\phi_{\,1}n_{1}\right)\exp\left(i\phi_{\,2}n_{2}\right)\\
&=\exp\left(i\phi_{\,0}n\right)\exp\left(i\phi_{+}\left(n_{1}+n_{2}\right)/2\right)\exp\left(i\phi_{-}\left(n_{1}-n_{2}\right)/2\right),
\end{split}
\end{equation}
where $n_{i}=a_{i}^{\dagger}a_{i}$, $n=n_{0}+n_{1}+n_{2}$, and $\phi_{\pm}=\left(\phi_{\,1}\pm\phi_{\,2}\right)$, $\phi_{\,0}=2Jt_{\omega}\cos\left(\theta/3+2\pi/3\right)$, $\phi_{\,1}=2\sqrt{3}Jt_{\omega}\sin\left(\theta/3\right)$, and $\phi_{\,2}=2\sqrt{3}Jt_{\omega}\sin\left(\theta/3+\pi/3\right)$. Actually, it is a hard work to obtain the unitary transformation of this system via multiplying that of each step, however, the approach we used here only needs the calculation of Eq.~(\ref{E9}) to get an equivalent unitary parametrization transformation Eq.~(\ref{E10}). Note that if the input state is a path-symmetric pure state~\cite{PhysRevA.85.011801,PhysRevA.93.033859}, a QFI formula for relative phase $\phi_{-}$ will be given by
\begin{equation}\label{E11}
F_{Q}(\phi_{-})=\Delta^{2}\left(n_{1}-n_{2}\right),
\end{equation}
and $\Delta\theta$ can be got via the error propagation formula
\begin{equation}\label{E12}
\Delta\theta=\sqrt{3}\Delta\phi_{-}/2Jt_{\omega}\cos\left(\theta/3-\pi/3\right).
 \end{equation}
\section{\label{sec:level3}OPTIMAL INPUT STATE}
In this section, we investigate the variance of Hermitian operator $\mathcal{H}$ with various input states under the condition of no particle loss. To compare different resources equivalently, we take into account the same average particle number $\bar{n}=N$ for each state.

\subsection{\label{sec:level4}Particle number state}

 First of all, we consider the particle number state which is very common as an input state in quantum metrology. Eq.~(\ref{E9}) shows that the Hermitian operator $\mathcal{H}$ is the function of $a^{\dagger}_{i}a_{i}$, and the precision of the phase will be more accurate with a larger variance of number operators. According to Refs.~\cite{PhysRevLett.102.040403,PhysRevA.80.013825}, a general input state for three-mode is expressed as $\left|\psi\right\rangle_{\rm in}=\sum_{m,n=0}^{m+n=N}c_{m,n}\left|m,n,N-m-n\right\rangle$ where $\sum_{m,n=0}^{m+n=N}\left|c_{m,n}\right|^{2}=1$.
 From Eq.~(\ref{E10}), one can see that the first term $e^{i\phi_{\,0}n}$ only provides a global phase and can be ignored. Furthermore, it is easy to find that the QFI only depends on the relative number of particles from two modes, which means concentrating particles in two modes to increase the variance of number operators is the optimal way to improve the QFI.
 
 In this article, we set site zero as a phase reference (phase remains constant), and only the phase changes of the other two modes are considered here. For the uncorrelated state, the BAT state involved in Ref.~\cite{PhysRevA.81.043624}, $F_{Q}$ for parameter $\phi_{-}$ are $N$ and $N\left(N/2+1\right)$, respectively. And for the NOON state, which is considered as the optimal state in the lossless case in Ref.~\cite{PhysRevA.81.043624}, gives the maximal QFI of $F_{Q}(\phi_{-})=N^{2}$ . Moreover, we find that  $\Delta\theta$ can be improved slightly by utilizing $\Delta\phi_{+}$ and the maximally entangled state $\left|\psi\right\rangle_{M}=1/\sqrt{2}\left(\left|N,N\right\rangle+\left|0,0\right\rangle\right)$ that also gives the maximal QFI of $F_{Q}(\phi_{+})=N^{2}$. Note that the error propagation
 formula of $\Delta\phi_{+}$ is $\Delta\theta=\Delta\phi_{+}/2Jt_{\omega}\cos\left(\theta/3+\pi/6\right)$, and the minimal uncertainty of the phase $\theta$ is given by
 \begin{equation}\label{E13}
 \Delta\theta_{min}=\frac{1}{2NJt_{\omega}}.
 \end{equation}
 In contrast with the precision given by the NOON state and Eq.~(\ref{E12}), $\Delta \theta$ can be improved by $\sqrt{3}$ times with $\phi_{+}$ and $\left|\psi\right\rangle_{M}$.
 
 \subsection{\label{sec:level5}Entangled coherent state}
 Entangled coherent state (ECS) presents the superiority  for phase estimation in optical interferometers~\cite{PhysRevLett.107.083601}, and its advantage still exists in the atomic gyroscope. In general, the ECS is given  by
 \begin{equation}\label{E14}
 \left|\psi\right\rangle_{E}=\mathcal{N}_{\alpha}\left(\left|\alpha,0\right\rangle+\left|0,\alpha\right\rangle\right)
 \end{equation}
 where $\left|\alpha\right\rangle=\exp(\alpha a^{\dagger}-\alpha^{*}a))\left|0\right\rangle$ is a coherent state and the normalization factor is $\mathcal{N}_{\alpha}=\left(2+2e^{-\left|\alpha\right|^{2}}\right)^{-1/2}$. The normalization factor must satisfy that $2\mathcal{N}_{\alpha}^{2}\left|\alpha\right|^{2}=N$ for the same average particle number $N$. On the situation of no particle loss, we make the approximation that $\left|\alpha\right|^{2}\approx N$ and $\mathcal{N}_{\alpha}\approx 1/\sqrt{2}$ when $N\gg1$. Note that if we use the parameter $\phi_{-}$ or Eq.~(\ref{E9}) to estimate the phase $\theta$, then we will obtain the QFI of $F_{Q}\approx N(N+1)$. This is indeed correct for two-mode input states in the absence of a phase reference~\cite{PhysRevA.85.011801,PhysRevA.93.033859}, but for the gyroscope we consider here, there is an                                                                           extra mode (i.e., site zero), which can be seen as a phase reference. It allows us to estimate $\phi_{\,1}$ or $\phi_{\,2}$ directly instead of the relative phase $\phi_{-}$ between different modes. Utilizing Eqs.~(\ref{E2}), (\ref{E3}) and (\ref{E10}), the maximal QFI is given by
 \begin{equation}\label{E15}
 F_{Q}\left(\phi_{\,1(2)}\right)= 
 4\Delta^2n_{1(2)}\approx N(N+2).
 \end{equation}
 This precision is consistent with the result obtained in Ref.~\cite{PhysRevLett.107.083601}. Hence, the minimal uncertainty of the phase $\theta$ is given by,
 \begin{equation}\label{E16}
 \Delta\theta_{min}\approx\frac{\sqrt{3}}{2\sqrt{N(N+2)}Jt_{\omega}}.
 \end{equation}
 Eq.~(\ref{E16}) shows that the ECS produces a better precision than the NOON state involved in Ref.~\cite{PhysRevA.81.043624}, especially when $N$ is modest. In addition, we can also use the parameter $\phi_{+}$ and a state similar to $\left|\psi\right\rangle_{M}$, that is, $\left|\psi\right\rangle_{EM}=\mathcal{N}_{\alpha}\left(\left|\alpha,\alpha\right\rangle+\left|0,0\right\rangle\right)$, to improve $\Delta\theta$ slightly. Finally, the minimal $\Delta\theta$ is given by,
 \begin{equation}\label{E17}
 \Delta\theta_{min}\approx\frac{1}{2\sqrt{N(N+1)}Jt_{\omega}}.
 \end{equation}
 The reason the ECS has such an advantage is that $\left|\psi\right\rangle_{E}$ can be rewritten as~\cite{PhysRevLett.107.083601}
 \begin{equation}\label{E18}
 \left|\psi\right\rangle_{E}=\mathcal{N}_{\alpha}e^{-\left|\alpha\right|^{2}/2}\sum_{n=0}^{\infty}\frac{\alpha^{n}}{\sqrt{n!}}\left(\left|n,0\right\rangle +\left|0,n\right\rangle \right),
 \end{equation}
 and it can be regarded as a NOON-like state, which is the superposition of the NOON states with different particle numbers, and the QFI is proportional to the square of the number of particles, thus the NOON states with larger particle numbers in Eq.~(\ref{E18}) would improve the precision significantly.
 \subsection{\label{sec:level6}Squeezed entangled state}
 
 Inspired by the ECS, it is easy to get an idea that we can turn the coherent state in the ECS into the squeezed vacuum state $\left|\xi\right\rangle=\exp\left[\left(\xi^{*}a^2-\xi(a^{\dagger})^{2}\right)/2\right]\left|0\right\rangle$ to obtain the squeezed entangled state (SES), that is,
 \begin{equation}\label{E19}
 \left|\psi\right\rangle_{S}=\mathcal{N}_{\xi}\left(\left|\xi,0\right\rangle+\left|0,\xi\right\rangle\right),
 \end{equation}
 where $\mathcal{N}_{\xi}=\left(2/\cosh r+2\right)^{-1/2}$, $r=\left|\xi\right|$. And $r$ is determined by $\cosh r/(1+\cosh r)\sinh^{2}r=N$ in the case of the fixed average particle number. The SES has been discussed in Refs.~\cite{PhysRevA.93.033859,Lee2016}, and a generation scheme is  proposed in Ref.~\cite{Lee2016}. Likewise, we can use $\phi_{\,1}$ or $\phi_{\,2}$ to obtain the variance of $\theta$. With a phase reference, the maximal QFI is $F_{Q}(\phi_{\,1(2)})\approx 5N^{2}+4N$ in the large squeezing regime $r\gg1$, which is almost 2 times better than the QFI of $F_{Q}\left(\phi_{-}\right)\approx 3N^{2}+2N$ with respect to the relative phase $\phi_{-}$ involved in Refs.~\cite{PhysRevA.93.033859,Lee2016}. And the minimal phase uncertainty $\Delta \theta$ is given by
 \begin{equation}\label{E20}
 \Delta\theta_{min}\approx\frac{\sqrt{3}}{2\sqrt{N(5N+4)}Jt_{\omega}}.
 \end{equation}
 Similarly, utilizing $\phi_{+}$ and $\left|\psi\right\rangle_{SM}=\mathcal{N}_{\xi}\left(\left|\xi,\xi\right\rangle+\left|0,0\right\rangle\right)$, we obtain a smaller phase uncertainty, that is
 \begin{equation}\label{E21}
 \Delta\theta_{min}\approx\frac{1}{2\sqrt{N(3N+2))}Jt_{\omega}},
 \end{equation}
 and we see that the precision is improved $\sqrt{3}$ times compared with the NOON state and the ECS.
 
\begin{figure}
 	\centering
 	\includegraphics[width=96mm,height=74mm]{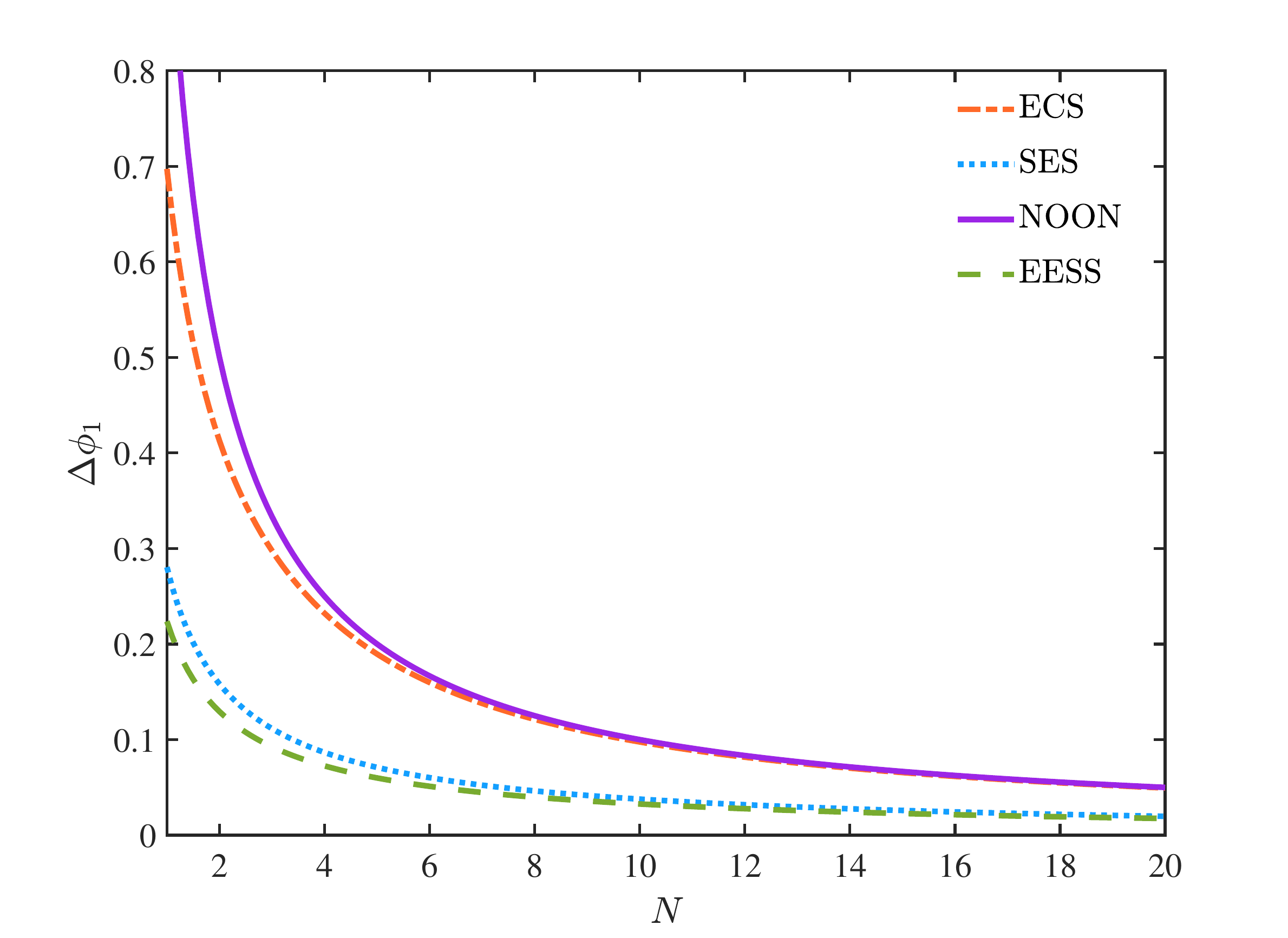}% Here is how to import EPS art
 	\caption{\label{fig:2}  Phase uncertainty $\Delta \phi_{\,1}$ varies with the number of particles in lossless case. $\Delta\phi_{\,1}$ of the NOON state (purple solid  line) is approximately equal to that of the ECS (red dot-dashed line) for large $N$; $\Delta\phi_{1}$ of the EESS (green dashed line) is minimal in the low particle-number regime, however, when the particle number increases, the precision of the EESS is roughly equal to that of the SES.   }
 \end{figure}

  Fig. \ref{fig:2} shows the phase uncertainty $\Delta \phi_{\,1}$ for various quantum states varies with the average particle number $N$ in the lossless case. We find that when the average particle number is fixed, the ECS is superior to the NOON state only if $N$ is small, however, if we choose the SES as the input state, the precision is enhanced significantly regardless of the number of particles. To appreciate this, Eq.~(\ref{E19}) is rewritten as
 \begin{equation}\label{E22}
 \left|\psi\right\rangle_{S}=\mathcal{N}_{\xi}\sum_{n=0}^{\infty}C_{2n}\left(\left|2n,0\right\rangle+\left|0,2n\right\rangle\right),
 \end{equation}
 where $C_{2n}=1/\sqrt{\cosh r}\left(-e^{i\vartheta}\rm{tanh}r/2\right)^{n}\sqrt{(2n)!}/n!$, $\vartheta=\rm arg(\xi)$. It is easy to find that SES is a coherent superposition of the NOON states with different even numbers. It means  larger particle number states are included in the SES than the ECS under the condition of the same average particle number, which is beneficial to improve the precision of the gyroscope. 
 
 Furthermore, Refs.~\cite{Lee2016,PhysRevA.93.033859,Rubio_2019} show that the QFI is proportional to Mandel $\mathcal{Q}$-parameter of the single-mode in path-symmetric state, and in this article the QFI for parameter $\phi_{\,1}$ can be rewiritten as
 \begin{equation}\label{E23}
 F_{Q}=2N(\mathcal{Q}+1),
 \end{equation}
 where $\mathcal{Q}=\Delta^2n_{1}/\left\langle n_{1}\right\rangle-1 $, hence the squeezed vacuum state has an advantage over the coherent state for its super-Poissonian statistics.
 
 \subsection{\label{sec:level7} Entangled even squeezed state}
 
 In the previous result, we reveal the relationship between the QFI and Mandel $\mathcal{Q}$-parameter, and one can immediately see that the states with larger Mandel $\mathcal{Q}$-parameters can give larger QFI. In general, the Mandel $\mathcal{Q}$ -parameter of the  even coherent state is larger in comparision with the coherent state, and we follow this idea to consider the superpostion of two squeezed vacuum states, i.e.,
 \begin{equation}\label{E24}
 \left|\Phi\right\rangle=\mathcal{N}\left(\left|\xi\right\rangle+
 \left|-\xi\right\rangle\right),
 \end{equation}  
 where $\mathcal{N}=\left(2+2\left(\cosh2r\right)^{-1\text{/2}}\right)^{-1/2},r=\left|\xi\right|$. Then we introduce the "entangled even squeezed state" (EESS) which is an original NOON-like state and constructed with $\left|\Phi\right\rangle$, that is
 \begin{equation}\label{E25}
 \left|\psi\right\rangle_{ES}=\mathcal{N}_{\Phi}\left(\left|\Phi,0\right\rangle
 +\left|0,\Phi\right\rangle\right)
 \end{equation} 
 where
 \begin{equation}\label{E26}
 \mathcal{N}_{\Phi}=\frac{\left(1+\left(\cosh2r\right)^{-1\text{/2}}\right)^{1/2}}{\left[2+2\left(\cosh2r\right)^{-1/2}+4\left(\cosh r\right)^{-1}\right]^{1/2}}.
 \end{equation} 
 Although $\left|\xi\right\rangle$ is an even state, the superposition shown in Eq.~(\ref{E24}) can generate a more "even" state. The analytical expression of the QFI for parameter $\phi_{\,1}$ is presented in Appendix B, and the phase uncertainty $\Delta\phi_{\,1}$ of EESS with respect to $N$ is shown in Fig \ref{fig:2}. We find the EESS has the best performance in the low-particle-number regime, but similar to the case of the NOON state and the ECS, the precision of the EESS and the SES tend to be the same as the average particle number increases. The specific explanation is presented in Appendix B. Then we set $\mathrm{arg}(\xi)=0$, and Eq.~(\ref{E25}) can be expanded with Fock states, that is
 \begin{equation}\label{E27}
 \left|\psi\right\rangle_{ES}=\mathcal{N}_{\xi,-\xi}\sum_{m=0}^{\infty}D_{4m}\left(\left|4m,0\right\rangle +\left|0,4m\right\rangle \right),
 \end{equation}
 where $D_{4m}=2\left(\cosh r\right)^{-1/2}\left(\tanh r/2\right)^{2m}\sqrt{\left(4m\right)!}/\left(2m\right)!$, $\mathcal{N}_{\xi,-\xi}=\mathcal{N}_{\Phi}\mathcal{N}$. Note that the EESS only involves number states that are multiples of four, and this leads to larger number states are included in the EESS than the SES for a fixed average particle number, which means larger QFI can be obtianed. Moreover, from Eq.(\ref{E23}) we see that the Mandel $\mathcal{Q}$-parameter of the single-mode component $\left|\Phi\right\rangle$ in the EESS determines the QFI directly. For a more specific example, taking $N=2$ and the $\mathcal{Q}$ for the SES is $\mathcal{Q}_{SES}\approx9$, and  $\mathcal{Q}_{EESS}\approx14$ for the EESS. In Ref.~\cite{Rubio_2019}, the Mandel $\mathcal{Q}$-parameter for various states are given, and we see that the EESS has an even larger $\mathcal{Q}$ than the squeezed cat state ($\mathcal{Q}\approx11.75$), which is considered as the optimal state in Ref.~\cite{PhysRevA.93.033859}. Additionally, $\phi_{+}$ and $\mathcal{N}_{\Phi}\left(\left|\Phi,\Phi\right\rangle+ \left|0,0\right\rangle\right)$ can be employed to enhance the precision slightly, and in the large squeezing regime
 we can get the same result shown in Eq.~(\ref{E20}).

\section{\label{sec:level8}EFFECTS OF PARTICLE LOSS}

In practice, decoherence and particle loss are inevitable, thus it is necessary to take into account the effects of particle loss of the atomic gyroscope in this section. In Refs.~\cite{PhysRevA.81.043624,PhysRevA.88.043832,PhysRevA.80.013825}, the related issues have been involved. In this paper, we still use the model that inserting two fictitious "beam splitter" with the transmission rate $\eta$ into two sites and, in general, the output state is described by a mixed state $\rho$. In this scheme, the particle loss is the loss from the momentum modes during $t_{\omega}$~\cite{PhysRevA.81.043624}. We assume that both modes have the same loss rates $R=1-\eta$, and the model in Ref.~\cite{PhysRevA.80.013825} is used here to describe the effect of particle loss on the QFI. The situations of the particle number state and the ECS have been discussed in Refs.~\cite{PhysRevA.80.013825,PhysRevLett.107.083601,PhysRevA.88.043832}, thus we focus on the calculation of the SES and the EESS in this article, and more 
specific calculations are presented in Appendix C. 

First of all, we should calculate the reduced density matrix of the SES after particle loss. The particle number states in Eq.~(\ref{E22}) evolve into
\begin{equation}\label{E28}
\begin{split}
\left|2n,0\right\rangle \rightarrow\sum_{l_{a}=0}^{2n}\sqrt{B_{l_{a}}^{2n}}\left|2n-l_{a},0\right\rangle \otimes\left|l_{a},0\right\rangle,\\
\left|0,2n\right\rangle \rightarrow\sum_{l_{b}=0}^{2n}\sqrt{B_{l_{b}}^{2n}}\left|0,2n-l_{b}\right\rangle \otimes\left|0,l_{b}\right\rangle,
\end{split}
\end{equation}
where
\begin{equation}\label{E29}
B_{l_{a(b)}}^{2n}=\left(\begin{array}{c}
2n\\
l_{a(b)}
\end{array}\right)\eta^{2n-l_{a(b)}}\left(1-\eta\right)^{l_{a(b)}},
\end{equation}
and $\left|l_{a},0\right\rangle$ and $\left|0,l_{b}\right\rangle$ are the states which represent $l_{a},l_{b}$ particles are lost from  site one and site two, respectively. Then we obtain the reduced density matrix which is expressed as
\begin{equation}\label{E30}
\rho=p_{0,0}\rho_{0,0}+\sum_{l_{a}=1}^{\infty}p_{l_{a},0}\rho_{l_{a},0}+\sum_{l_{b}=1}^{\infty}p_{0,l_{b}}\rho_{0,l_{b}},
\end{equation}
where $\rho_{l,m}=\left|\psi_{l,m}\right\rangle\left\langle\psi_{l,m}\right|$, and $l,m$ are the number of particles lost. When $l=m=0$, we have
\begin{equation}\label{E31}
\left|\psi_{0,0}\right\rangle=\frac{1}{\sqrt{p_{0,0}}}\mathcal{N}_{\xi}\sum_{n=0}^{\infty}C_{2n}\eta^{n}\left(\left|2n,0\right\rangle +\left|0,2n\right\rangle \right).
\end{equation}

\begin{figure}
	\centering\includegraphics[width=94mm,height=72mm]{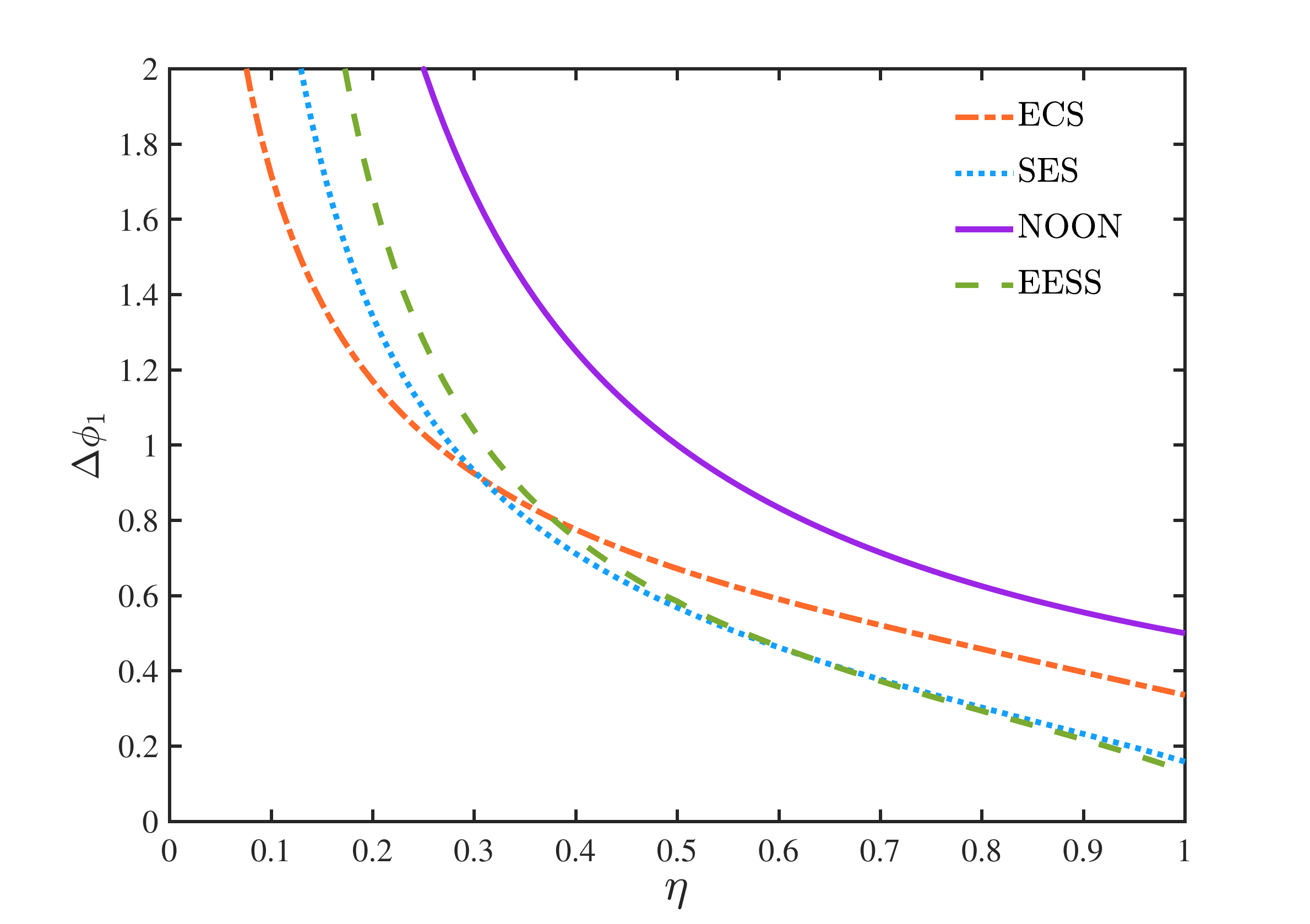}% Here is how to import EPS art
	\caption{\label{fig:3}  Phase uncertainty $\Delta \phi_{\,1}$ varies with transmission rate $\eta$ for $N=2$. The green dashed line and the blue dotted line show the precision of $\phi_{\,1}$ for the EESS and the SES, respectively. The ECS and the NOON state which are investigated in the previous studies are depicted in
		red dot-dashed line and purple solid line for comparison.  }
\end{figure}

While $l=l_{a},m=0$ and $l=0,m=l_{b}$, we also obtain
\begin{equation}\label{E32}
\begin{split}
\left|\psi_{l_{a},0}\right\rangle=\frac{1}{\sqrt{p_{l_{a},0}}}\mathcal{N}_{\xi}\left(\sum_{n=\Gamma}^{\infty}C_{2n}\sqrt{B_{l_{a}}^{2n}}\left|2n-l_{a},0\right\rangle \right),\\
\left|\psi_{0,l_{b}}\right\rangle =\frac{1}{\sqrt{p_{0,l_{b}}}}\mathcal{N}_{\xi}\left(\sum_{n=\Gamma}^{\infty}C_{2n}\sqrt{B_{l_{b}}^{2n}}\left|0,2n-l_{b}\right\rangle \right),
\end{split}
\end{equation}
where
\begin{equation}\label{E33}
\Gamma=\begin{cases}
\left(l_{a(b)}+1\right)/2 & {\rm for}\ l_{a(b)}\rm \ is\ odd,\\
\ l_{a(b)}/2 & {\rm for}\ l_{a(b)}\rm \ is\ even.
\end{cases}
\end{equation}
$p_{0,0}$, $p_{l_{a},0}$ and $p_{0,l_{b}}$ are the normalization factors required for calculation of the mixed state $\rho$. We set $\vartheta=0$, $\xi=r$, then Eq.~(\ref{E29}) can be rewritten as
\begin{equation}\label{E34}
\left|\psi_{0,0}\right\rangle=\mathcal{N}_{\tilde{\xi}}\left(\left|\tilde{r},0\right\rangle +\left|0,\tilde{r}\right\rangle \right),
\end{equation}
where $\tilde{r}=\mathrm{arc}\tanh\left(\eta\tanh r\right)$, and $\mathcal{N}_{\tilde{\xi}}=\left(2/\cosh \tilde{r}+2\right)^{-1/2}$. Utilizing Eqs.~(\ref{E31}), (\ref{E32}), (\ref{E34}) and the normalized property of states, we can obtain
\begin{equation}\label{E35}
\begin{split}
&p_{0,0}=\frac{1+\cosh\tilde{r}}{1+\cosh r},\\
\sum_{l_{a}=1}^{\infty}&p_{l_{a},0}=\sum_{l_{b}=1}^{\infty}p_{0,l_{b}}=
\frac{\cosh r-\cosh\tilde{r}}{2\left(1+\cosh r\right)}.
\end{split}
\end{equation}

To obtain the QFI of the mixed state $\rho$, we have to diagonalize the mixed state as $\rho=\sum_{m}\lambda_{m}\left|\lambda_{m}\right\rangle\left\langle\lambda_{m}\right|$, where $\lambda_{m}$ is an eigenvalue and $\left|\lambda_{m}\right\rangle$ is eigenvector. Here we first estimate the phase $\theta$ through $\phi_{\,1}$, then choose average particle number $N=2$, and then truncate the particle number state at $n=56$ in Eq.~(\ref{E31}). For a mixed state $\rho$, the QFI is expressed as \cite{Liu_2014,LIU2014167,Yu:18}
\begin{equation}\label{E36}
F_{Q}=4\sum_{m}\lambda_{m}\left\langle\lambda_{m}\right|n_{1}^2\left|\lambda_{m}\right\rangle-\sum_{m,m'}
\frac{8\lambda_{m}\lambda_{m'}}{\lambda_{m}+\lambda_{m'}}\left|\left\langle\lambda_{m}\right|n_{1}\left|\lambda_{m'}
\right\rangle\right|^{2},
\end{equation}
and the numerical simulation of QFI is shown in Fig. \ref{fig:3}. Here we give the specific calculation of the SES, and the case of the EESS is very similar so we put it into Appendix C.

In Fig. \ref{fig:3} we present the variation of $\Delta \phi_{\,1}$ with the change of $\eta$. In the low loss regime, the EESS gives the best precision compared with the SES state, the ECS, and the NOON state. This advantage in precision persists until $\eta<0.5$, and is inferior to the ECS when $\eta<0.37$. In general, the loss rates in an experiment is smaller than $50\%$ \cite{PhysRevA.81.043624}, and thus the EESS is the preferred state to obtain the maximum precision. 

\section{\label{sec:level9}CONCLUSION}
In this paper, we have investigated the optimal input state in the atomic gyroscope which is based on three-site Bose-Hubbard model, and the main approach is to find the state that has the maximal QFI. We have provided a Hermitian operator $\mathcal{H}$  which contains the dynamical information of the whole process, hence the measurement procedure can be simplified as an equivalent unitary transformation. To obtain the maximal QFI, we take the squeezed entangled state (SES) and the entangled even squeezed state (EESS) as candidates to improve the precision of the atomic gyroscope.

Compared with the particle number state (especially the NOON state) and the entangled coherent state, the best precision is achieved by taking the EESS as the input state in the ideal case. In addition, the existence of an extra mode allows us to set up a phase reference, which can make it possible to estimate the phase change of a mode directly. And on this basis, we found that using another form of entangled states, such as $\left|\psi\right\rangle_{M}$, $\left|\psi\right\rangle_{EM}$ and $\left|\psi\right\rangle_{SM}$ can obtain slightly better precision in the lossless case. Moreover, the QFI under practical conditions is also considered, and we found that the EESS improve the precision significantly in moderate loss regime. Although we face some challenges that there is no current experimental scheme or technology is proposed to generate the EESS, it is still the preferred input state in this measurement scheme. And the results we obtained in this article also fully demonstrate the EESS has a great potential in quantum metrology. Furthermore, we believe that the method we used in this article is helpful to the people who want to improve the precision of other measurement systems.

\begin{acknowledgments}
 This work was supported by the National Key Research and Development Program of China (Grants No. 2017YFA0304202
and No. 2017YFA0205700), the NSFC (Grant No. 11875231 and
No. 11935012), and the Fundamental Research Funds for the Central Universities through Grant No. 2018FZA3005.
\end{acknowledgments}
\appendix

\section{THE DERIVATION OF $\mathcal{H}$}\label{Appendix:A}
First, we know that step (\romannumeral2) to (\romannumeral6) can be regarded as unitary transformations, and we use $\left\lbrace U_{2}...U_{6}\right\rbrace $ to represent step (\romannumeral2)-(\romannumeral6). For a pure state, the QFI with respect to parameter $\theta$ is given by
\begin{equation}\label{A1}
F_{Q}=4\left(\left\langle\partial_{\theta}\psi|\partial_{\theta}\psi\right\rangle-\left|\left\langle\psi|\partial_{\theta}\psi\right\rangle\right|^{2}\right).
\end{equation}

In step (\romannumeral1) an initial state $\left|\psi\right\rangle_{\rm in}$ is prepared, and $\left|\psi\right\rangle=U_{6}U_{5}...U_{2}\left|\psi\right\rangle_{\rm in}=\prod_{i=2}^{6}U_{8-i}\left|\psi\right\rangle_{\rm in}$. Note that only $U_{4}$ is the function of $\theta$, and utilizing Eq.~(\ref{E2}) the QFI with respect with $\theta$ is express as 
\begin{equation}\label{A2}
\begin{split}
F_{Q}&= 4\left(_{\rm in}\!\left\langle\psi\right|U_{2}^{\dagger}U_{3}^{\dagger}\mathcal{H}_{4}^{2}U_{3}U_{2}\left|\psi\right\rangle_{\rm in}-\left|_{\rm in}\!\left\langle\psi\right|U_{2}^{\dagger}U_{3}^{\dagger}\mathcal{H}_{4}U_{3}U_{2}\left|\psi\right\rangle_{\rm in}\right|^{2}\right)\\
&=4 \left(_{\rm in}\!\left\langle\psi\right|\mathcal{H}^{2}\left|\psi\right\rangle_{\rm in}-\left|_{\rm in}\!\left\langle\psi\right|\mathcal{H}\left|\psi\right\rangle_{\rm in}\right|^{2}\right)
\end{split}
\end{equation}
where $U_{4}=e^{-iH_{r}t_{\omega}/\hbar}$,$\mathcal{H}_{4}=i\left(\partial_{\theta}U_{4}^{\dagger}\right)U_{4}=-\partial_{\theta}H_{r}t_{\omega}/\hbar$, and the equality $\left(\partial_{\theta}U^{\dagger}\right)U=-U^{\dagger}\left(\partial_{\theta}U\right)$ is used here. The specific expression of $\mathcal{H}_{4}$ is given by 
\begin{equation}\label{A3}
\begin{split}
\mathcal{H}_{4}&=\frac{-2Jt_{\omega}}{3}\left[\sin\left(\frac{\theta+2\pi}{3}\right)n_{-1}^{\alpha}+\sin\left(\frac{\theta}{3}\!\right)n_{0}^{\alpha}+\sin\left(\frac{\theta-2\pi}{3}\right)n_{1}^{\alpha}\right]\\
&=\frac{iJt_{\omega}}{3}e^{\frac{i\theta}{3}}\left(a_{0}^{\dagger}a_{1}
+a_{1}^{\dagger}a_{2}+a_{2}^{\dagger}a_{0}\right)+h.c.,
\end{split}
\end{equation}
where $n_{j}^{\alpha}=\alpha_{j}^{\dagger}\alpha_{j}$, $j=-1,0,1$, and Eq.~(\ref{E5}) is used to obtain the second line of Eq.~(\ref{A3}).
In order to obtain $\mathcal{H}$, we have 
\begin{equation}\label{A4}
U_{3}^{\dagger}\mathcal{H}_{4}U_{3}=\frac{iJt_{\omega}}{3}e^{\frac{i\theta}{3}}\left(a_{0}^{\dagger}a_{1}
+a_{1}^{\dagger}a_{2}e^{-i\frac{4\pi}{3}}+a_{2}^{\dagger}a_{0}e^{i\frac{4\pi}{3}}\right)+h.c.
\end{equation}
Then we use quasi-momentum basis in the following calculation, and Eq.~(\ref{A4}) becomes
\begin{equation}\label{A5}
U_{3}^{\dagger}\mathcal{H}_{4}U_{3}=\frac{iJt_{\omega}}{3}e^{\frac{i\theta}{3}}\left(\alpha_{-1}^{\dagger}\alpha_{0}+\alpha_{0}^{\dagger}\alpha_{1}e^{-i\frac{2\pi}{3}}+\alpha_{1}^{\dagger}\alpha_{-1}e^{i\frac{2\pi}{3}}\right)+h.c.,
\end{equation}
and $U_{2}$ in Eq.~(\ref{E7}) is expressed as 
\begin{equation}\label{A6}
U_{2}=e^{i\frac{2\pi}{9}\left(2\alpha_{0}^{\dagger}\alpha_{0}-\alpha_{-1}^{\dagger}\alpha_{-1}-\alpha_{1}^{\dagger}\alpha_{1}\right)}.
\end{equation}
Utilizing Eqs.~(\ref{A5}) and (\ref{A6}), we have
\begin{equation}\label{A7}
\begin{split}
\mathcal{H}&=U_{2}^{\dagger}U_{3}^{\dagger}\mathcal{H}_{4}U_{3}U_{2}\\
&=\frac{iJt_{\omega}}{3}e^{i\frac{\left(\theta+2\pi\right)}{3}}\left(\alpha_{-1}^{\dagger}\alpha_{0}+\alpha_{0}^{\dagger}\alpha_{1}+\alpha_{1}^{\dagger}\alpha_{-1}\right)+h.c.
\end{split}
\end{equation}
Finally, we utilize the initial basis $\left(a_{0},a_{1},a_{2}\right)^{\rm T}$ and obtain 
Eq.~(\ref{E9})
\begin{equation}\label{A8}
\mathcal{H}=-\frac{2Jt_{\omega}}{3}\left[\sin\left(\frac{\theta+2\pi}{3}\right)n_{0}+\sin\left(\frac{\theta\!-2\pi}{3}\right)n_{1}+\sin\left(\frac{\theta}{3}\right)n_{2}\right].
\end{equation}

\section{THE ANALYTICAL EXPRESSION OF THE QFI OF THE ESES}\label{Appendix:B}
To obtain the QFI of the EESS with respect to parameter $\phi_{\,1}$, we need to calculate $\Delta^2 n_{1}$. For a squeezed vacuum state $\left|\xi\right\rangle=S(\xi)\left|0\right\rangle$, we have
\begin{equation}\label{B1}
\left\langle\xi|-\xi\right\rangle=\frac{1}{\sqrt{\cosh2r}},
\end{equation}
\begin{equation}\label{B2}
\left\langle\xi|a^{\dagger}a|-\xi\right\rangle
=-\frac{\sinh^{2}r}{\left(\cosh2r\right)^{3/2}},  
\end{equation}
\begin{equation}\label{B3}
\left\langle \xi\right|a^{\dagger}aa^{\dagger}a\left|-\xi\right\rangle=-\frac{\left(3+\cosh2r\right)\sinh^{2}r}{2\left(\cosh2r\right)^{5/2}}
\end{equation}
and here we set $\rm arg(\xi)=0$. It is known that $\Delta^{2} n_{1}=_{ES}\!\!\left\langle\psi|n_{1}^{2}|\psi\right\rangle\!_{ES}-\left|_{ES}\!\left\langle\psi|n_{1}|\psi\right\rangle\!_{ES}\right|^{2}$, and the first term of the right side is
\begin{equation}\label{B4}
\left\langle n_{1}^2\right\rangle=2\mathcal{N}_{\xi,-\xi}^{2}\left(2\sinh^{2}r\cosh^{2}r+\!\sinh^{4}r-\frac{\left(3+\cosh2r\right)\sinh^{2}r}{2\left(\cosh2r\right)^{5/2}}\right),
\end{equation}
and the second term is 
\begin{equation}\label{B5}
\begin{split}
\left\langle n_{1}\right\rangle^2&=4\mathcal{N}_{\xi,-\xi}^{4}\sinh^{4}r\left(1-\left(\cosh2r\right)^{-3/2}\right)^{2}\\
&=\frac{N^2}{4}
\end{split}
\end{equation}
where $\mathcal{N}_{\xi,-\xi}=\left(1+(\cosh 2r)^{-1/2}+2(\cosh r)^{-1}\right)^{-1/2}/2$. The second line of Eq.~(\ref{B5}) is because the total average particle number is $N$. Finally, the QFI of the EESS is give by
\begin{equation}\label{B6}
\begin{split}
F_{Q}(\phi_{\,1})&=\frac{2N\left(2\cosh^{2}r+\sinh^{2}r-\frac{\left(3+\cosh2r\right)}{2\left(\cosh2r\right)^{5/2}}\right)}{1-\left(\cosh2r\right)^{-3/2}}-N^{2}\\
&=N\left(\frac{3+3(\cosh2r)^2+\cosh2r}{\cosh2r}+\frac{3\left(\cosh2r\right)^{1/2}}{1+\left(\cosh2r\right)^{1/2}+\cosh2r}-N\right)
\end{split}
\end{equation}

In the large squeezing regime $r\gg1$, we have $\mathcal{N}_{\xi,-\xi}\approx1/2$, $\sinh^{2}r=(\cosh2r-1)/2\approx N$, and $F_{Q}$ becomes
\begin{equation}\label{B7}
F_{Q}(\phi_{\,1})\approx 5N^2+4N.
\end{equation}
This result shows that the QFI of the EESS is approximately equal to that of the SES under condition of large $N$.
\section{PARTICLE LOSS OF THE ESES}\label{Appendix:C}
In this Appendix, we give a specific approach to calculate the effect of particle loss of the EESS. The reduced density matrix of the EESS is expressed as 
\begin{equation}\label{C1}
\bar{\rho}=\bar{p}_{0,0}\bar{\rho}_{0,0}+\sum_{l_{a}=1}^{\infty}\bar{p}_{l_{a},0}\bar{\rho}_{l_{a},0}+\sum_{l_{b}=1}^{\infty}\bar{p}_{0,l_{b}}\bar{\rho}_{0,l_{b}},
\end{equation}
where $\bar{\rho}_{l,m}=\left|\bar{\psi}_{l,m}\right\rangle\left\langle\bar{\psi}_{l,m}\right|$, and $l,m$ are the number of particles lost.  Similar to  Eqs.~(\ref{E31}) and (\ref{E32}), we have
\begin{equation}\label{C2}
\left|\bar{\psi}_{0,0}\right\rangle=\frac{1}{\sqrt{\bar{p}_{0,0}}}\mathcal{N}_{\xi,-\xi}\sum_{n=0}^{\infty}D_{4m}\eta^{2m}\left(\left|4m,0\right\rangle +\left|0,4m\right\rangle \right),
\end{equation}
and
\begin{equation}\label{C3}
\begin{split}
\left|\bar{\psi}_{l_{a},0}\right\rangle =\frac{1}{\sqrt{\bar{p}_{l_{a},0}}}\mathcal{N}_{\xi,-\xi}\left(\sum_{\{m|4m\geq l_{a},m\in z\}}\!\!\!\!\!\!\!\!\!\!\!\!\!\!D_{4m}\sqrt{B_{l_{a}}^{4m}}\left|4m-l_{a},0\right\rangle \right),\\
\left|\bar{\psi}_{0,l_{b}}\right\rangle =\frac{1}{\sqrt{\bar{p}_{0,l_{b}}}}\mathcal{N}_{\xi,-\xi}\left(\sum_{\{m|4m\geq l_{b},m\in z\}}\!\!\!\!\!\!\!\!\!\!\!\!\!\!D_{4m}\sqrt{B_{l_{b}}^{4m}}\left|0,4m-l_{b}\right\rangle \right),
\end{split}
\end{equation}
where $\bar{p}_{0,0}$, $\bar{p}_{l_{a},0}$ and $\bar{p}_{0,l_{b}}$ are the normalization factors. To obtain these three factors, $D_{4m}\eta^{2m}$ in Eq.~(\ref{C2}) can be written as
\begin{equation}\label{C4}
\begin{split}
D_{4m}\eta^{2m}&=\frac{2}{\sqrt{\cosh r}}\left(\frac{1}{2}\eta\tanh r\right)^{2m}\frac{\sqrt{(4m)!}}{(2m)!}\\
&=\sqrt{\frac{\cosh\tilde{r}}{\cosh r}}\frac{2}{\sqrt{\cosh\tilde{r}}}\left(\frac{1}{2}\tanh\tilde{r}\right)^{2m}\frac{\sqrt{(4m)!}}{(2m)!}\\
&=\sqrt{\frac{\cosh\tilde{r}}{\cosh r}}\widetilde{D}_{4m},
\end{split}
\end{equation}
where $\tilde{r}=\rm arctanh(\eta\tanh r)$. Then we rewrite Eq.~(\ref{C2}) as
\begin{equation}\label{C5}
\begin{split}
\left|\bar{\psi}_{0,0}\right\rangle&=\frac{1}{\sqrt{\bar{p}_{0,0}}}\mathcal{N}_{\xi,-\xi}\sum_{n=0}^{\infty}D_{4m}\eta^{2m}\left(\left|4m,0\right\rangle +\left|0,4m\right\rangle \right)\\
&=\frac{1}{\sqrt{\bar{p}_{0,0}}}\sqrt{\frac{\cosh\tilde{r}}{\cosh r}}\frac{\mathcal{N}_{\Phi}\mathcal{N}}{\widetilde{\mathcal{N}}}\sum_{m=0}^{\infty}\widetilde{D}_{4m}\widetilde{\mathcal{N}}\left(\left|4m,0\right\rangle +\left|0,4m\right\rangle\right)\\
&=\mathcal{N}_{\tilde{\Phi}}\left(\left|\tilde{\Phi},0\right\rangle +\left|0,\tilde{\Phi}\right\rangle \right)
\end{split}
\end{equation}
where $\widetilde{\mathcal{N}}=\left(2+2\left(\cosh2\tilde{r}\right)^{-1\text{/2}}\right)^{-1/2}$, $\left|\tilde{\Phi}\right\rangle=\mathcal{N}\left(\left|\tilde{r}\right\rangle+
\left|-\tilde{r}\right\rangle\right)$, and
\begin{equation}\label{C6}
\mathcal{N}_{\tilde{\Phi}}=\frac{\left(1+\left(\cosh2\tilde{r}\right)^{-1/2}\right)^{1/2}}{\left[2+2\left(\cosh2\tilde{r}\right)^{-1/2}+4\left(\cosh \tilde{r}\right)^{-1}\right]^{1/2}}.
\end{equation} 
Note that we set $\arg(\xi)=0$ here. From Eqs.~(\ref{C5}) and (\ref{C6}), we can obtain
\begin{equation}
\bar{p}_{0,0}=\frac{\mathcal{N}^{2}}{\mathcal{\widetilde{N}}^{2}}\frac{\cosh\tilde{r}}{\cosh r}\frac{\mathcal{N}_{\Phi}^{2}}{\mathcal{N}_{\tilde{\Phi}}^{2}}.
\end{equation}
For $\bar{p}_{l_{a},0}$ and $\bar{p}_{0,l_{b}}$, we have
\begin{equation}
\begin{split}
\sum_{l_{a}=1}^{\infty}\bar{p}_{l_{a},0}&=\sum_{l_{b}=1}^{\infty}\bar{p}_{0,l_{b}}\\
&={\sum_{l_{a}=0}^{\infty}}\mathcal{N}_{\Phi}^{2}\mathcal{N}^{2}\!\!\!\!\!\!\!\!\!\!\!\!\!\sum_{\{m|4m\geq l_{a},m\in z\}}\!\!\!\!\!\!\!\!\!\!\!\!\!D_{4m}^{2}B_{l_{a}}^{4m}-\mathcal{N}_{\Phi}^{2}\mathcal{N}^{2}\sum_{m=0}^{\infty}D_{4m}^{2}B_{0}^{4m}\\
&=\mathcal{N}_{\Phi}^{2}\left(1-\frac{\cosh\tilde{r}}{\cosh r}\frac{\mathcal{N}^{2}}{\mathcal{\widetilde{N}}^{2}}\right).
\end{split}
\end{equation}
In fact, $\bar{p}_{l,m}$ is the probability of the density matrix with the corresponding particle loss. And utilizing the relationship between $\mathcal{N}_{\Phi}$ and $\mathcal{N}_{\tilde{\Phi}}$
\begin{equation}
\frac{\mathcal{N}^{2}\cosh\tilde{r}}{\mathcal{\widetilde{N}}^{2}\cosh r}\left(\frac{1-2\mathcal{N}_{\tilde{\Phi}}^{2}}{\mathcal{N}_{\tilde{\Phi}}^{2}}\right)=\frac{1-2\mathcal{N}_{\Phi}^{2}}{\mathcal{N}_{\Phi}^{2}},
\end{equation}
It's easy to verify that  $p_{0,0}+\sum_{l_{a}=1}^{\infty}\bar{p}_{l_{a},0}+\sum_{l_{b}=1}^{\infty}\bar{p}_{0,l_{b}}=1$.

\bibliography{mybib}
\end{document}